\documentclass{elsart}

\usepackage{harvard}

\usepackage{graphicx}

\usepackage{amssymb}

\def\url#1{{\ttfamily\def\/{/\discretionary{}{}{}}#1}}

\begin{document}

\begin{frontmatter}
\title{Simulations of Nonthermal Electron Transport in Multidimensional
Flows: Application to Radio Galaxies}


\author[Minn]{T.W. Jones\thanksref{tj}}, 
\author[Minn]{I.L. Tregillis\thanksref{it}} \&
\author[Ryu]{Dongsu Ryu\thanksref{dr}}

\thanks[tj]{E-mail: twj@astro.spa.umn.edu}
\thanks[it]{E-mail: tregilli@msi.umn.edu}
\thanks[dr]{E-mail: ryu@canopus.chungnam.ac.kr}

\address[Minn]{School of Physics and Astronomy,  University of Minnesota, 116 Church St. S.E., Minneapolis, MN 55455}
\address[Ryu]{Department of Astronomy and Space Science, Chungnam National University, Daejeon, 305-764, Korea}

\begin{abstract}

We have developed an economical, effective numerical scheme for cosmic-ray
transport suitable for treatment of electrons up to a few
hundreds of GeV in multidimensional simulations of radio galaxies.
The method follows the electron population in sufficient detail to allow 
computation of synthetic radio and X-ray observations of the 
simulated sources, including spectral properties (see the companion
paper by Tregillis \etal ~1999). The cosmic-ray particle simulations can
follow the effects of shock acceleration, 
second-order Fermi acceleration as well as radiative and
adiabatic energy losses. We have applied this scheme to 
2-D and 3-D MHD simulations of jet-driven flows
and have begun to explore links between dynamics and
the properties of high energy electron populations in radio lobes.
The key initial discovery is the great importance to the high
energy particle population of the very unsteady and inhomogeneous flows,
especially near the end of the jet. Because of this,
in particular, our simulations show that a large fraction of the particle 
population flowing from the jet into the cocoon never passes through strong shocks.
The shock strengths encountered are not simply predicted by 1-D models, and
are quite varied.
Consequently, the emergent electron spectra are highly heterogeneous. Rates
of synchrotron aging in ``hot-spots" seem similarly to be very
uneven, enhancing complexity in the spectral properties
of electrons as they emerge into the lobes and making more difficult the task
of comparing dynamical and radiative ages.
\end{abstract}

\end{frontmatter}

\section{Introduction}
\label{intro}

The jet-based, dynamical paradigm for radio galaxies now seems secure, but
our understanding of the basic physics remains primitive. 
A key barrier has been our limited ability to fill the
enormous gap between complex plasma flows thought to
drive the phenomenon and the emissions that
reveal them. The emissions
come from particles very far from thermodynamic equilibrium,
so that their energy distributions depend on local microphysics and their histories. 
Further, the posited plasma flows are strongly
driven systems, so they are also mostly far from any stable dynamical
equilibria. Among other things this leads to very unsteady flows.
These characteristics make numerical simulations
a necessity to push beyond the simplest characterizations
of the phenomena and to verify even simple analytical
calculations based on equilibrium assumptions. 

The past decade
has seen a revolution in our ability to carry out sophisticated
multidimensional gasdynamical simulations, including magnetic fields
(e.g., Clarke \etal~1989; Lind \etal~1989; K\"ossl \etal~1990; Nishikawa \etal~1998).
The diffusive shock acceleration paradigm for particle energization,
which is relatively robust in relating bulk dynamics and particle
spectra,
seems to offer an attractive way past ignorance of some microphysical
details. Despite these advances,
modeling the relevant particle microphysics adequately in simulations
has remained illusive. 
These difficulties come from both severe technical constraints on 
simulations and ignorance of some of the most important physical parameters.
Now we have initiated a program that takes
a significant technical step forward, and which we hope will help resolve
some of the key physical ambiguities. In multi-dimensional MHD flow simulations
we follow the nonthermal electron population in some detail,
allowing us to compute for the first time meaningful model emissions.
This paper
describes some initial results as they apply to the properties
of the nonthermal electrons. A companion paper (Tregillis
\etal ~1999) describes application of our methods to synthetic radio and X-ray observations
and what they seem to be telling us about interpretation of observations
in terms of physical
source parameters. A report of initial 2-D simulations, plus
additional references are given in Jones \etal~1999.
The 3-D results shown here are preliminary. A full report is in
preparation.

\section{Background}
\label{methods}

Our flow dynamics is treated with a second-order accurate, conservative, ``TVD''
ideal MHD code described in Ryu \& Jones 1995; Ryu \etal~1995; Ryu \etal~1998.
It maintains the divergence free condition to the magnetic field to
machine accuracy using an upwinded constrained transport scheme as
described in Ryu \etal~1998.
For energetic particle transport
we use the conventional ``convection diffusion equation''
for the momentum distribution function, $f$, (e.g., Skilling 1975)
which follows spatial and momentum diffusion as well as
spatial and momentum advection of the particles. 
The last of these corresponds to
energy losses and gains from, for example, adiabatic expansion and
synchrotron aging.

However, high computational costs 
prohibit solving this equation through standard finite difference methods in complex
flows expected in radio galaxies. To circumvent this
we use a conservative finite volume approach in the momentum coordinate, taking advantage
of the broad spectral character expected for $f(p)$. 
Particle fluxes across momentum bin boundaries are estimated
by representing $f(p)$ as $f(p) \propto p^{-q(p)}$, where $q(p)$ varies in a 
regular way.
Numerically we use the integrated number of electrons within each bin
and the slope, $q$, within each bin.
Thus, we can follow electron spectral evolution in smooth flows for all the effects
mentioned above with a modest number of
momentum bins. Typically we have used 8 bins to cover energies up to a few 
hundred GeV for electrons. We have shown that this approach 
produces solutions in good agreement with more conventional
methods, including diffusive acceleration at shocks and
synchrotron energy losses. However, in the
flows being studied, diffusive acceleration of electrons to GeV energies
at shocks is effectively instantaneous within a dynamical time step.
Because of that direct simulation of this physics would be 
prohibitively expensive.
We can, however, also circumvent this difficulty if we assume the analytic, steady,
test particle form for the electron
distribution just behind shocks. Ignoring for this discussion some 
details, that spectrum will be a
power law with an index, $q = \frac{3 r}{r - 1}$, where $r$ is the shock
compression ratio. 

Together these features give us a 
powerful tool for numerical simulations of such complex phenomena as 
radio galaxies (Jones \etal~1999; Jun \& Jones 1999). 
Our method is complementary to that recently described by Micono \etal~1999.
Those authors followed the convection-diffusion equation in detail outside 
of shocks using conventional methods, but were constrained by the numerical
limitations we describe to follow a very limited number of Lagrangian
volume elements within the flow. That approach does not allow one to 
make synthetic observations of the simulated flows, whereas ours does.

\section{Discussion}
\label{discussion}

\begin{figure}
\begin{center}
\includegraphics*[width=10cm, angle=-90]{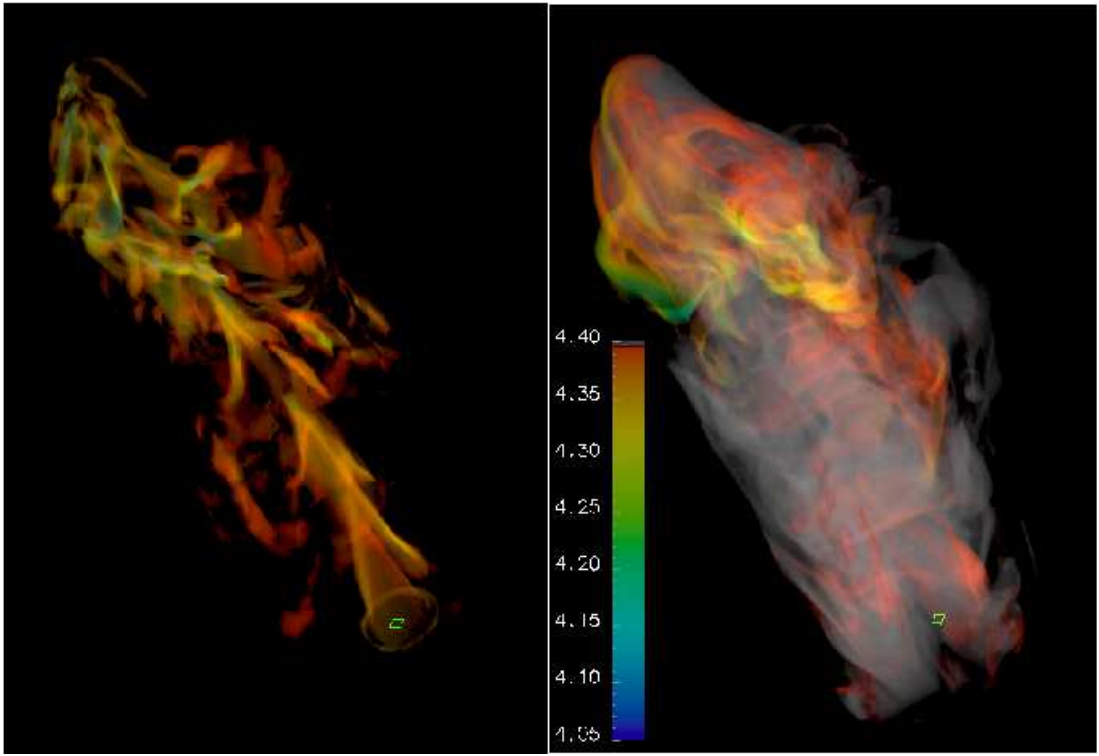}
\end{center}
\caption{Left: Volume rendering of $\nabla\cdot{\bf u}$, showing the
3-D distribution of shocks in a precessing jet flow $\sim$40 jet radii long.
Stronger shocks are bluer. Only jet plasma flow is shown. Right: The
distribution of 10 GeV electron momentum spectral indices, $q$,
as defined in the text. Spectra unchanged during propagation from the
jet origin are rendered in white. Shock-modified spectra are given
a color related to the degree of flattening of the spectrum. 
Bluer spectra are flatter. The jet origin
is marked by the green boxes in both images.}
\label{fig_jet}
\end{figure}

Our initial radio galaxy explorations with this new tool have focussed on improved
understanding of the ways that complex jet flow dynamics
influence the spectral properties of electrons.
For this purpose we carried out {2-D} axisymmetric
and fully {3-D} MHD simulations of light, pressure-balanced, supersonic jets 
carrying weak, but ``active'', magnetic fields, which are helical inside the
jet and axial in the uniform ambient medium. So far we have followed the
dynamics until the jets propagate $\sim 40~-~50$ jet radii, considering
cases where the magnetic fields are strong enough to cause significant 
synchrotron aging and cases where it is too weak for this to matter.
We have considered cases in which the relativistic electrons are
injected naturally as part of diffusive acceleration physics at shocks and
cases where only an electron population introduced with the jet flow is
represented. For the latter cases, we assume this population comes onto
the grid with a power law spectrum, $q = 4.4$, corresponding to a synchrotron
spectral index, $\alpha = \frac{q-3}{2} = 0.7$. In all cases the spectra evolve appropriately within the flows in response to shock acceleration, synchrotron 
aging and adiabatic effects.

Fig. 1 illustrates some of the important common properties in our
simulations. This case is a 3-D, Mach 8 (internal Mach number), 
light, MHD jet ($\rho_j/\rho_a = 10^{-2}$,
$\beta = p_b/p_g \approx 10^2$, where $\beta$ refers to the magnetic field
on the jet axis.), which was slowly
precessed on a 5 degree cone at the origin. The center of the jet 
origin is marked in the figure by the small
green boxes. 
This numerical experiment was carried out on a $576\times192\times192$
grid. The inflowing jet had a top hat velocity profile with a thin
transition sheath around it. The core radius spanned 15 zones.

This flow is not at all steady, since the jet terminus
tends to ``whip'' around, sometimes forming ``splatter spots'' (Williams \& Gull 1985)
and then pinching off and redirecting itself.
Here we show properties at a moment when the flow is relatively simple.
The left panel reveals through volume rendering of 
$\nabla\cdot {\bf u}$ the locations of shocks in the
jet and its back-flow. The strongest shocks appear blue, while
relatively weaker shocks are red. The right panel shows the spatial distribution
of the spectral index of $\sim 10$ GeV electrons at the same time. 
Synchrotron aging is negligible in this case, so we are seeing in the
electron spectra only effects from shock acceleration and advective mixing.
Shock-modified spectra flatter than the ``injected
spectrum'' are shown in color, with the relatively flatter (steeper) spectra being
blue (red). The injected spectral index, $q = 4.4$,  is rendered in white.
Given the absence of synchrotron aging, all spectra are at least as flat as that
entering at the jet origin.
For this simulation there were no additional electrons injected at 
shocks; that is, the entire relativistic electron population entered with the jet.
In this case those electrons constitute a fraction $10^{-4}$ of the
total number of electrons. The remainder are thermal. Since the
jet speed is assumed to be 0.1 c, thermal electrons in the jet have
energies $\sim 70$ keV.
Only flow regions filled with plasma originating in the jet are shown
in the figure.
So, for example, the bow shock preceding the jet is not visible in the
left panel, since it occurs in the ambient medium.

These images illustrate dramatically
that shock structures in jet driven flows are very complex, 
and also show our consequent key finding that the spectral properties of
electrons emerging into the cocoon are extremely heterogeneous.
Comparison of the images shows us why this finding makes sense.
First, let us locate the jet flow itself. Near its base the jet can be followed
through its conical internal shocks. The terminus of the jet occurs at the
far upper left in a small, strong shock. It turns out that only the central core
of the jet actually exits through that terminal shock.
Much of the plasma emerging from the jet has passed through only weak shocks
before it is redirected into the cocoon.
Consequently, shocks inside and at the end of the jet have had a relatively
small influence on the spectra of electrons entering the cocoon.
Note next that most of the rather complex ``shock web'' near the jet head 
involves cocoon flow, in fact. Remarkably, the strongest shocks are often 
in the cocoon, rather than the jet. They appear to be generated by the non-axial 
motions of the jet head. 

The distribution of electron spectra in the adjacent image does not map in an
obvious way onto the shock distribution. Some insight into this
complication comes from noting the apparent ``streams'' in the
particle spectra, which are especially apparent near the
jet terminus. The streams highlight flows downstream of localized,
strong shocks. This feature emphasizes that the shocks are themselves very complex,
and also that we are seeing a blend of many different flow histories.
The relative importance of shocks inside the cocoon compared to the
jet can be recognized
by understanding the origin of a relatively flat spectral region
visible in the figure roughly 2.5 cm to 4.5 cm from the end of the 
jet towards the origin (green box). There, $q \approx 4.3$, so it shows
yellow in the image. Examination of flow streaklines shows that this
plasma all passed through a small shock visible in the shock image
(with a yellow color) about 2.5 cm from the
end of the jet, but physically in the cocoon, not the jet itself.
These results emphasizing the complexity of the evolution of
particle spectra in jet head regions support and augment our earlier 
findings from 2-D axisymmetric simulations (Jones \etal ~1999). 




An analogous finding from the 2-D study  was that magnetic field structures
are highly inhomogeneous (``intermittent''),
especially in the head region. Thus, synchrotron emissivity and associated
aging rates are very inhomogeneous. That behavior seems even stronger in
our initial examination of a 3-D run that was similar to the one shown in Fig. 1,
except that the magnetic field was strong enough to produce significant
aging on the time scales simulated. Observational properties of these
flows are described in the companion paper by Tregillis \etal~1999. 

This work was supported at the University of Minnesota by the NSF through
grant AST96-16964 and by the University of Minnesota Supercomputing
Institute. DR was supported in part by KOSEF through grant 975-0200-06-2.


{\bf Q: D. Harris} Great code, but deductions about particle vs B field
distributions may be misleading, because you still have MHD, where basic
fluid controls where and how much e \& B occur/get carried. Is it not
true that in real jets the dynamics may be controlled by the relativistic
particles and B field?

{\bf A: T. Jones} Magnetic fields are fully dynamically coupled in our 
simulations. In fact, although we described the magnetic fields as ``weak''
because the nominal ratio of gas to magnetic pressure in the jet is large,
we find that there are numerous places in the cocoon where magnetic,
Maxwell stresses are significant. They are properly accounted for.
The relativistic particles represent a high energy tail to the bulk
population, so their dynamics accurately reflects the total. We have
restricted their energy content so that it does not dominate in order
that the overall flows may be treated non-relativistically. The
relativistic particles are handled correctly in the context of the
nonrelativistic bulk flow. Eventually we will want to do simulations 
like these with fully relativistic MHD codes. That step is some
distance off, even if we accept the good progress being made on
relativistic fluid codes. Treatment of particle acceleration at relativistic
shocks is much more subtle than at nonrelativistic shocks, so it is
not obvious how we will make that step for complex problems. 
Here we have assumed that the shocks are not relativistic and not superluminal.

\end{document}